\documentclass[12pt,preprint]{aastex}
\usepackage{epsfig}
\usepackage{bm}

\slugcomment{KSUPT-04/3 \hspace{0.5truecm} May 2004}

\shorttitle{Magnetic field and the CMB}
\begin{document}
\title{Cosmological Magnetic Fields vs. CMB}

\author{Tina Kahniashvili\altaffilmark{1,2}}
\altaffiltext{1}{Department of Physics, Kansas State University, 116
                 Cardwell Hall, Manhattan, KS 66506.}
\altaffiltext{2}{Center for Plasma Astrophysics, Abastumani Astrophysical
                 Observatory, A.~Kazbegi ave 2a, Tbilisi, 380060,
                 Republic of Georgia.}

\begin{abstract}
I present a short review of the effects of a cosmological
magnetic field on the CMB temperature and polarization anisotropies.
Various possibilities for constraining the magnetic field amplitude
are discussed.
\end{abstract}

Cosmological primordial seed magnetic fields were proposed to
explain the existence of observed $\sim 10^{-6}$G magnetic fields
in galaxies and clusters (see, e.g.,  Widrow 2002, Giovaninni 2003,
and references therein). To preserve approximate spatial isotropy
a seed vector magnetic field has to be small and hence can be
treated as a first order term in perturbation theory. If the
energy density parameter of a primordial magnetic field,
$\Omega_B = B^2/(4\pi \rho_{cr})$ (where $\rho_{cr}$ is the
critical density), is five or six  of order of magnitude less
than that of the radiation (photons),  $\Omega_B \sim
10^{-6}-10^{-5} \Omega_\gamma$, this is still of the  order of
the current accuracy of CMB measurements (Bennett et al. 2003),
so we might expect that such a field strength
($B \sim 10^{-8}-10^{-9}$Gauss) could leave detectable traces on CMB
temperature or polarization anisotropies.

Primordial magnetic fields could be generated during early epochs
of the Universe, such as during inflation, or the
electroweak phase transition, or might be generated by
primordial turbulence (for reviews see Grasso and Rubinstein 2001,
Widrow 2002, Giovaninni 2003). Cosmological magnetic fields induce
scalar (density), vector (vorticity)
and tensor (gravitational waves) fluctuations,
and through them influence the CMB
temperature and polarization anisotropies
(see Mack et al. 2002 and references therein).
Hence precise CMB measurements
(Bennett et al. 2003) can be used to  constrain
primordial magnetic fields. An interesting
possibility is to consider the rotation of the CMB
polarization plane due to the  Faraday effect
(Kosowsky and Loeb 1996).

The simplest illustrative case to consider is a homogeneous magnetic field
(Giovaninni and Shaposhnikov 1998), which generates
magnetosonic and Alfv\'en waves.
Due to the rescaling of sound velocity in a cosmological model with a
homogeneous magnetic field:
$c_S^2 \rightarrow c_S^2 + v_A^2$ (where
$v_A=B/\sqrt{4\pi(\rho+p)}$ is the Alfv\'en speed),
the influence of fast magnetosonic wave propogation on CMB anisotropies
consists of shifts in the accoustic  peak positions  (Adams et al. 1996).

In stardard cosmology vector perturbations decay with time  and so
do not affect the CMB. The presence of a homogeneous
magnetic field alters this
situation: such  a field supports Alfv\'en
(vorticity) waves, and also breaks
spatial statistical isotropy.
A homogeneous magnetic field hence induces non-zero off-diagonal correlations
between  temperature multipole coefficients.
In particular,  non-zero correlations between
$l$ and $l\pm2$ multipole coefficients
are given by (Durrer et al. 1998)
\begin{equation}
   D_l(m) = \langle a_{l-1,m}^* a_{l+1, m} \rangle =
   \langle a_{l+1,m}^* a_{l-1, m} \rangle .
   \label{dlm}
\end{equation}
Here the power spectrum
$D_l(m)$ depends on the primordial vorticity perturbation spectrum
(which we assume to be given by a simple power
law $P_\Omega(k) \propto k^n/k_D^{n+3}$), and the Alfv\'en speed $v_A$.
The presence of a non-zero $D_l(m)$ has a simple physical explanation:
the temperature anisotropy correlation between two points on the sky
depends not only on the angular separation between the two points but also
on their orientation with respect to the magnetic field.

An observational test to detect (or constrain) the presence of a
homogeneous cosmological magnetic field is based on computing the $D_l$
spectrum of CMB anisotropy data. Chen et al. (2004)
use the WMAP data to
constrain the magnetic field amplitude (at
illustrative  value of vorticity spectral index
$n=-7$ and $n=-5$) to be less than
about $10^{-8}-10^{-9}$ Gauss at three standard deviation.

A more realistic case\footnote{For cosmological  magnetic field
generation mechanisms see, e.g., Turner and Widrow (1988), Carroll and Field
(1991), Ratra (1992), Dolgov and Silk (1993), and Enqvist and Olesen (1994)}
to consider is a stochastic magnetic field with a
(Gaussian random) two-point correlation spectrum (Pogosian
et al. 2002):
\begin{equation}
\langle B^\star_i({\mathbf k})B_j({\mathbf k'})\rangle
=(2\pi)^3 \delta({\mathbf k}-{\mathbf k'}) [P_{ij} P_B(k)   +
i \epsilon_{ijl} \hat{k}_l P_H(k)]~,
\label{spectrum}
\end{equation}
where $P_B(k)$ ($\propto \langle|{\mathbf B}|^2 \rangle $) and
$P_H(k)$ ($\propto \langle {\mathbf B}
\cdot (\nabla \times {\mathbf B}) \rangle$)
are the symmetric and helical
magnetic field power spectra, respectively
(we assume that both are given by simple power laws), the plane
projector $P_{ij}\equiv \delta_{ij}-\hat{k}_i\hat{k}_j$,
$\epsilon_{ijl}$ is the totally antisymmetric tensor, and
$\hat{k}_i=k_i/k$.
The possibility of generating
helical magnetic fields
is discussed in
Vachaspati (2001) and Sigl (2002).
The symmetric part of the magnetic field in eq.(\ref{spectrum})
contributes to the CMB temperature  and polarization anisotropies
via induced vector and tensor perturbations
(for details see Mack et al. 2002)\footnote{
CMB temperature and polarization anisotropy vector mode
contributions
for a magnetic field spectrum peaked at a fixed value of $k$ are  given in
Subramanian and Barrow (1998), Seshadri and Subramanian (2001),
and Subramanian et al. (2003).
CMB temperature anisotropy induced by gravitational
waves generated by a magnetic field are discussed  in
Durrer et al. (2000)}. $P_B(k)$ induces
parity-even CMB fluctuations, with the following
maximum rate of growth with respect of $l$
\begin{equation}
C_l^{\theta \theta (V)} \propto l^2, ~~~~~~~
C_l^{EE (V)} \propto l^2, ~~~~~~~ C_l^{BB (V)} \propto l^2
~~~~~~~
C_l^{\theta E (V)} \propto l^2
\end{equation}
For a vector perturbation
the $BB$-power spectrum is slightly larger than the $EE$ one,
whereas the $EE$
and $\theta E$ power spectra are approximately equal
{\footnote{Temperature anisotropies are dominated by the vector dipole term,
which correlates poorly with the radial function corresponding
to $E$ polarization (Hu and White 1997), so the $\theta E$ power
spectrum is dominated by a subdominant temperature contribution arising
from the vector quadropole term, which then coincidentally
renders the spectrum in a form approximately identical to the
$E$ polarization power spectrum
itself.}}While $n_S \rightarrow -3$ corresponds to the symmetric 
magnetic field power spectrum being scale-invariant,  the CMB vector
power spectra are not flat for this value.

For tensor pertubations, the parity-even CMB power spectra generated
from the symmetric magnetic field
power spectrum are (Durrer et al. 2000,
Mack et al. 2002):
\begin{equation}
C_l^{\theta \theta (T)} \propto l, ~~~~~~~
C_l^{EE (T)} \propto l, ~~~~~~~~C_l^{BB(T)} \propto l,
~~~~~~~
C_l^{\theta E (T)} \propto l
\end{equation}
For magnetic field induced gravitational wave contribution to the CMB
anisotropies,
the $E$ polarization power spectrum is slightly larger than the $B$ one.
For $n_S > -3/2$
the polarization power spectra are comparable to the
temperature power spectra.
This is due to the fact that
both the temperature and polarization fluctuations are dominated by
the intrinsic temperature quadropole moments, which arise
from the gravitational wave solution
${\dot h}$ instead of being induced via free streaming
dipoles as in the case of a vector perturbations.
Also, for $n_S>-3/2$ the magnetic source term for the tensor mode is
approximately independent of $k$ and the resulting power spectra have the
well
known behaviour for a white noise source: $C_l l^2 \propto l^3$. As expected,
for a scale-invariant magnetic field with $n_S \rightarrow -3$,
the tensor part of the CMB power spectra is flat. Note
that our analytical approximations are valid for $l<500$ for the vector mode
and for $l<100$ for the tensor mode (due to the damping of gravitational
waves
when they enter horizon at decoupling), see Caprini et al.
(2004) for detailed discussion. Comparison with the WMAP CMB data
(Bennett et. al. 2003)
constrains the magnetic field amplitude to be less than about
 $10^{-9}$ Gauss
{\footnote{Constraints of a similar magnitude result from considering
resonant photon-graviton
conversion (Chen 1995), and from the distortion of the
CMB (Jedamzik et al. 2000)}}.

A magnetic field with non-zero helicity ($P_H(k)$) will
induce additional effects (Pogosian et al. 2002).
In particular, the presence of a helical part results in
non-zero parity-odd CMB power spectra, such as $C_l^{EB}$ and
$C_l^{\theta B}$.
Also, a  helical
magnetic field will generate gravitational waves with parity
odd spectra (Caprini et al. 2004).
Using the linear polarization basis, $e^T_{ij}=({\bf e_1} \times {\bf e_1} -
{\bf e_2} \times {\bf e_2})_{ij}$,
$e^\times_{ij}=({\bf e_1} \times {\bf e_2} +
{\bf e_2} \times {\bf e_1})_{ij}$) 
the helical part of the magnetic tensor source $g(k)$
can be directly connected with gravitational waves $h^T$ and $h^\times$
($h_{ij}=e^{T}_{ij}h^T+e^{\times}_{ij}h^\times$)
correlations:
\begin{equation}
\langle h^{\star T}({\bf k}) h^\times({\bf k^\prime})-
h^{\star T}({\bf k}) h^\times({\bf k^\prime}) \rangle
\propto i \delta({\bf k} -
{\bf k^\prime}) g(k)
\end{equation}
A magnetic field with helicity will also induce non-decaying vorticity waves
  (Pogosian et al. 2002).
Both modes (vector and tensor) generate CMB temperature
and polarization anisotropies.
The helical part contributions to parity-even total CMB power spectra
are negative, but due to the causality restriction, $P_B(k) >|P_S(k)|$
and $n_S\leq n_A$, the total $C_l$'s are positive. The ratio between the
helical and symmetric part contributions to the parity-even CMB
power spectra $C_{l,H}/C_{l, S}$ depends on the corresponding indices
$n_H$ and $n_S$, as well as on $P_H(k)$ and $P_B(k)$.
The parity-odd power spectra are generated by $P_H(k)$, but
are dependent on both the  spectral amplitude and index.
(For the tensor mode see Caprini et al. (2004) and for the
vector mode a paper is in preparation).

As  mentioned above, the presence of a cosmological magnetic field
results also
in the rotation of the CMB polarization plane via the Faraday effect
(Kosowsky and Loeb 1996).
Assuming that the rotational effect on polarization generated by the
magnetic field itself is a second order effect, and also that only scalar
perturbations are present, Faraday rotation
will generates B-polarization. In a current project (Kosowsky et al. 2004)
we study the Faraday rotation effect (and resulting
B-polarization signal) due to a
stochastic magnetic field. We show that an average rotation measure
is independent of  the helical part of magnetic field. 
Hence  a measurement of the   
rotation measure can constrain the symmetric part of the
magnetic field spectrum. The resulting $B$ polarization depends on the
initial polarization spectrum ($C_{l, in}^{EE}$) and on the rotation
angle power spectrum ($C_l^{\alpha \alpha}$) as (Kosowsky et al. 2004):
\begin{eqnarray}
\langle a_{l'm'}^{B\,*}{}'a_{lm}^B{}'\rangle =
\delta_{ll'}\delta_{mm'}N_l^2 \sum_{l_1l_2}&&
N_{l_2}^2 K(l,l_1,l_2)^2 C^{EE}_{l_2,in}C^{\alpha \alpha}_{l_1}
{(2l_1+1)(2l_2+1)\over 4\pi(2l+1)}
\nonumber\\&&
\left(C^{l0}_{l_10l_20}\right)^2
\label{answer}
\end{eqnarray}
where
\begin{equation}
K(l,l_1,l_2)\equiv -{1\over 2}\left(L^2 + L_1^2 + L_2^2 -2L_1L_2
-2L_1L +2L_1-2L_2 -2L\right),
\end{equation}
with $L=l(l+1)$,
$N_l\equiv (2(l-2)!/(l+2)!)^{1/2}$, and
$C^{l0}_{l_10l_20}$ are Clebsh-Gordon coefficients.

Assuming precise measurements of CMB temperature and polarization
anisotropies, Faraday rotation allow us to reconstruct the symmetric
magnetic field
spectrum, and since the total CMB power spectra depend
on both the symmetric and the helical parts of the magnetic field spectrum,
we can also constrain magnetic helicity.
Also, there is the theoretical possibility of reconstructing
magnetic helicity from the magnetic-field-generated gravitational wave
spectrum.

Our conclusions are as follows:

A homogeneous magnetic field, via generated  Alfv\'en waves,
induces non-zero off-diagonal correlations of CMB anisotropies 
multipoles coefficients.
The magnetic field can thus be constrained  by testing CMB data
for non-gaussianity.

A helical magnetic field generates gravitational waves
with parity odd spectra. This could serve, in principle, as a  method
for the  detection of 
the helicity of the magnetic field.

The Faraday rotation measurement can'not constrain
magnetic helicity. Thus, only the symmetric part of magnetic field
spectrum can  be reconstructed from RM measurements.

Current CMB data constrains the amplitude of a cosmological magnetic field 
to be less than order $10^{-9}$ Gauss.

{\underline{Acknowledgements}\\
The author thanks the Dark Matter 2004 conference organising committee
for hospitality.
The author acknowledges her collaborators, C.~Caprini, G.~Chen, R.~Durrer,
P.~Ferreira, A.~Kosowsky, G.~Lavrelashvili, A.~Mack, P.~Mukherjee,
B.~Ratra, Y.~Wang, A.~Yates, and thanks P.~Chen and
T.~Vachaspati for discussions. This work is supported by
NSF CAREER grant AST-9875031, DOE
EPSCoR grant DE-FG02-00ER45824, and CRDF-GRDF grant 3316.


\begin{thebibliography}{}
\bibitem{adams96}
  Adams, J., Danielsson, U.~H., Grasso, D., \and Rubinstein, H.,
  Phys.~Lett.~B, 388, 253 (1996)

\bibitem{ben03}Bennett, C. L., et al., Ap. \ J. \ S., {\bf 148}, 1
(2003)

\bibitem{capr04} Caprini, C., Durrer, R.,  \and Kahniashvili, T.,
Phys. \ Rev. \ D. {\bf 69}, 063006 (2004)

\bibitem{carroll91} Carroll, S.M., \and Field, G.B.,
Phys.\ Rev.\ D {\bf 43}, 3789 (1991).

\bibitem{chen04} Chen, G., Mukherjee, P., Kahniashvili, T., Ratra, B.,
\and Wang, Y., astro-ph/0403695, Ap. \ J. in press (2004)

\bibitem{chen95} Chen, P., Phys. \ Rev. \ Lett. {\bf 74}, 634 (1995)

\bibitem{dolgov93} Dolgov, A.D., \and Silk, J.,
Phys.\ Rev.\ D {\bf 47}, 3144 (1993)

\bibitem{durr98}Durrer, R., Kahniashvili, T.,
\and Yates, A., Phys. \ Rev. \ D. {\bf 58}, 123004 (1998)

\bibitem{durr00} Durrer, R. Ferreira, P., \and Kahniashvili, T.,
Phys. \ Rev. \ D. {\bf 61} 043001, (2000)

\bibitem{enqvist94} Enqvist, K., \and Olesen, P.,
Phys.\ Lett.\ B {\bf 329}, 195 (1994).

\bibitem{giov03}Giovannini, M., astro-ph/0312614 (2003)

\bibitem{giov98}Giovannini, M., \and Shaposhnikov, M.,
Phys. \ Rev. \ Lett., {\bf 80}, 22 (1998)

\bibitem{hu97} Hu, W. \and White, M., Phys. \ Rev. \ D.
{\bf 56 } 596 (1997)

\bibitem{jedamzik00} Jedamzik, J., Katalini\'c, V., \and Olinto, A.V.,
Phys.\ Rev.\ Lett.\ {\bf 85}, 700 (2000).

\bibitem{kosowsky96} Kosowsky, A., \and Loeb, A.,
Ap.\ J. {\bf 469}, 1 (1996).

\bibitem{kos04} Kosowsky, A., Kahniashvili, T., Lavrelashvili G., \and
Ratra, B.,  in preparation, (2004)

\bibitem{mack02} Mack, A., Kahniashvili, T., \and  Kosowsky, A.,
Phys. \ Rev. \ D. {\bf 65}, 123004 (2002)

\bibitem{pogosian02}Pogosian, L., Vachaspati, T.,
\and Winitzki, S., Phys. \ Rev. \ D. {\bf 65}, 083502 (2002)

\bibitem{ratra92}Ratra, B.,~Ap. \ J. {\bf 391}, L1 (1992)

\bibitem{seshadri01} Seshadri, T.~R., \and  Subramanian, K.,
Phys. \ Rev. \ D. {\bf 87}, 101301 (2001)

\bibitem{sigl02} Sigl, G.,
Phys.\ Rev. \ D. {\bf 66}, 123002 (2002)

\bibitem{subramanian03} Subramanian, K., Seshadri, T.~R.,
\and Barrow, J.~D., MNRAS, {\bf 344}, L31 (2003)

\bibitem{turner88} Turner, M.S., and Widrow, L.M.,
Phys.\ Rev.\ D {\bf 37}, 2743 (1988)

\bibitem{vachaspati91} Vachaspati, T.,
Phys.\ Lett.\ B {\bf 265}, 258 (1991)

\bibitem{vachaspati01} Vachaspati, T.,
Phys.\ Rev. \ Lett. {\bf 87}, 251302 (2001)

\bibitem{widrow03} Widrow, L.~A., Rev.~Mod.~Phys., {\bf 74}, 775 (2002)
\end{thebibliography}
\end{document}